\documentclass[preprint]{aastex}  

\slugcomment{Submitted to The Astronomical Journal}

\shorttitle{NIRSPEC spectra of Terzan~4 and Terzan~5}
\shortauthors{Origlia \& Rich}

\begin{document}
\title{High resolution infrared spectra of Bulge Globular Clusters: \\
The extreme chemical abundances of Terzan~4 and Terzan~5}

\author{Livia Origlia}
\affil{Osservatorio Astronomico di Bologna, Via Ranzani 1,
I--40127 Bologna, Italy}
\email{origlia@bo.astro.it}

\author{R. Michael Rich}
\affil{Math-Sciences 8979,
Department of Physics and Astronomy, University of California
at Los Angeles, Los Angeles, CA 90095-1562}
\email{rmr@astro.ucla.edu}

\altaffiltext{1}{Data presented herein were obtained
at the W.M.Keck Observatory, which is operated as a scientific partnership
among the California Institute of Technology, the University of California,
and the National Aeronautics and Space Administration.
The Observatory was made possible by the generous financial support of the
W.M. Keck Foundation.}

\begin{abstract}

Using the NIRSPEC spectrograph at Keck II, we have obtained
infrared echelle spectra covering the range $1.5-1.8~\mu \rm m$ for
the highly reddened bulge globular clusters Terzan~4 and
Terzan~5.
The clusters are of interest because a blue
horizontal branch in Terzan~4 is consistent with very low metallicity,
while Terzan~5 has been proposed as possibly the most metal rich
known Galactic globular cluster.
We report the first detailed abundance analysis
for stars in these clusters, and we find
[Fe/H]=--1.60 and --0.21~dex for Terzan~4 and Terzan~5, respectively,
confirming the presence of a large metallicity spread
in the bulge as well as in the halo of the Galaxy.
We find $\rm [\alpha/Fe]\approx+0.5$~dex in
Terzan~4 and $\rm [\alpha/Fe]\approx +0.3$~dex  in Terzan~5, consistent
with what has been found for field stars in the bulge.  The enhanced
$\alpha-$element abundances are also consistent with rapid
chemical enrichment, most likely by type~II SNe.
Finally, we measure
very low  $^{12}$C/$^{13}$C$\le10$ ratios
in all the giant stars, confirming that
extra-mixing mechanisms due to {\it cool bottom processing}
occur during evolution along the RGB.

\end{abstract}


\section{Introduction}

The detailed study of the abundance patterns of iron-peak, CNO and $\alpha$ elements
in field and globular cluster stellar populations is a fundamental step
required to understand
the star formation and chemical evolution history of our Galaxy.
These elements are synthesized in stars of different masses,
hence on different timescales and play a crucial role in disentangling
evolutionary mixing effects from primordial enrichment
(Wheeler, Sneden \& Truran 1989, McWilliam 1997)

Globular clusters represent the best astrophysical templates for the
study of the early Galactic evolution since they are the oldest
objects and their age is well constrained,
allowing a reliable calibration of a possible age--metallicity
relation (McWilliam 1997), a fundamental constraint for any chemical evolution model.
In addition, they are equally present in
the halo, the thick disk and the bulge of our Galaxy
(see e.g. Armandroff 1989, Minniti 1995),
therefore they represent unique chemical fossils to trace the early epoch of
the Galaxy formation.

Reliable elemental abundances are now available for
field stars (see e.g.  Boesgaard et al. 1999; Gratton et al. 2000,
Carretta, Gratton \& Sneden 2000 and reference therein)
while are still very incomplete for globular clusters.
In halo/disk field stars the average [$\alpha$/Fe] abundance ratio
shows a general enhancement of 0.3--0.5 dex with respect to the Solar
value up to [Fe/H]$\approx$--1 and a linear decreasing trend towards
Solar [$\alpha$/Fe] with further increasing metallicity.
The actual position of the knee (i.e. the
metallicity at which [$\alpha$/Fe] begins to decrease) depends on the
type~Ia SN timescales and it is also a function
of the Star Formation (SF) rate, while the amount of
$\alpha $ enhancement depends on the initial mass function of the progenitors
of the type~II SNe (see McWilliam 1997).

While the classic halo globular clusters have a mean
[$\alpha$/Fe]$\approx$+0.4, too few measurements have been
done at the high metallicity end to define the trend there
(e.g. Kraft 1994, Carney 1996).
Bulge globular clusters are ideal targets to study the behavior
of the abundance patterns in the high metallicity domain,
but foreground extinction is so great
as to largely preclude optical studies of any kind, particularly
at high spectral resolution.

The most accurate abundance determinations
in the Galactic bulge obtained so far and based on
high resolution optical spectroscopy refer
to a sample of K giants in Baade's window
(McWilliam \& Rich 1994, hereafter MWR94;
Rich \& McWilliam 2000)
and a few giants in two globular clusters, NGC~6553 and NGC~6528
(Barbuy et al. 1999; Cohen et al. 1999; Carretta et al. 2001).

With the availability of NIRSPEC, a high throughput infrared (IR) echelle
spectrograph at the Keck Observatory (McLean et al. 1998), came the prospect
of measuring the composition of the clusters in the bulge.
Recently, we observed
two bright giants in Liller~1 and NGC~6553 with NIRSPEC
and our abundance analysis has been presented in Origlia, Rich \& Castro
(2002, hereafter ORC02).
We find $\rm [Fe/H]=-0.3\pm0.2$ and $\rm [O/Fe]=+0.3\pm 0.1$
We also measure strong lines for the other $\alpha-$elements
Mg, Ca, and Si, obtaining an overall $\rm [\alpha/Fe]=+0.3\pm0.2$ dex.

In this paper we present the high resolution
IR observations and the abundance analysis for
a sample of bright giants in two additional globular clusters,
namely Terzan~4 and Terzan~5.   These clusters are of special
interest because they span the entire abundance range present
in the bulge.
These clusters suffer from large foreground extinction
(E(B--V)$\ge$2), precluding any optical study at high spectral resolution.
Recent HST--NICMOS photometry (Ortolani et al. 2001)
suggests old ages for these clusters
and confirms previous photometric estimates for their metallicities
(Ortolani, Barbuy \& Bica 1996, 1997):
roughly Solar for Terzan~5 and as low as 1/100 Solar for Terzan~4.
Ortolani et al. (1997) argue that a blue horizontal
branch is present in Terzan~4,  however the small numbers
of stars in their color-magnitude diagram (CMD) requires an independent
metallicity measurement in order for the case to be more convincing.
More recent NICMOS photometry of Terzan~5 (Cohn et al.
2002) confirms the presence of an old turnoff point and a CMD
consistent with high metallicity.  The recent
photometry confirms that both clusters lie within $\sim 1$ kpc of
the Galactic center.

Our observations and data reduction follow in Sect.~2.
Sect.~3 discusses our abundance analysis and
in Sect.~4 the resulting metallicities and radial velocities are presented.
Our concluding remarks are given in Sect.~5.

\section{Observations and Data Reduction}

Near infrared, high-resolution echelle spectra of four bright giants
in the cores of the bulge globular
clusters Terzan~4 and Terzan~5 have been acquired
on 23 June 2001 and 15 July 2002.
We used the infrared spectrograph NIRSPEC (McLean et al. 1998)
which is at the Nasmyth focus of the Keck~II telescope.
The high resolution echelle mode, with a slit width of $0\farcs43$
(3 pixels) and a length of 24\arcsec\
and the standard NIRSPEC-5 setting, which
covers most of the 1.5--1.8 micron H-band,
have been selected.  Typical exposure times (on source)
ranged from 4 to 16 minutes.
Fig.~1 shows the H band images of the observed core region of
Terzan~4 and Terzan~5
taken with the slit viewing camera
(SCAM) of NIRSPEC, which
has a field of view of 46\arcsec$\times$46\arcsec\
and a scale of
$0\farcs183$$~pixel^{-1}$.

The raw spectra of the observed stars have been reduced using the
REDSPEC IDL-based package written
at the UCLA IR Laboratory.
Each order has been
sky subtracted by using nodding pairs and flat-field corrected.
Wavelength calibration has been performed using arc lamps and a second order
polynomial solution, while telluric features have been removed by
dividing by the featureless spectrum of an O star.
At the NIRSPEC resolution of R=25,000 several single
roto-vibrational OH lines and
CO bandheads can be measured to derive accurate oxygen and carbon abundances.
Other metal abundances can be derived from the atomic lines
of Fe~I, Mg~I, Si~I, Ti~I and Ca~I.
Abundance analysis is performed by using full spectral synthesis
techniques and equivalent width measurements of representative lines.

\section{Abundance Analysis}

We compute suitable synthetic spectra
of giant stars by varying the stellar parameters and the
element abundances using an updated
version of the code described in Origlia, Moorwood \& Oliva (1993)
(see also ORC02).
The code uses the LTE approximation and is based
on the molecular blanketed model atmospheres of
Johnson, Bernat \& Krupp (1980)
which have been extensively used for abundance
studies based on high resolution spectra of cool stars
(e.g. Lambert et al. 1984, Smith \& Lambert 1985, 1990).
For temperatures above 4000~K the code uses the ATLAS9 models.
Since in the IR the major source of continuum opacity is H$^-$
with its minimum near 1.6 $\mu$m,
the dependence of the results on the choice of model
atmospheres is much less critical than in the visual range.

Three main compilations of
atomic oscillator strengths are used:
the Kurucz database
(c.f. {\it http://cfa-www.harward.edu/amdata/ampdata/kurucz23/sekur.html}),
Bi\`emont \& Grevesse (1973)
and Mel\'endez \& Barbuy (1999, hereafter MB99).
On average, MB99 log-gf values are systematically lower than Kurucz ones but
in most cases the difference does not exceed 0.2~dex
and the overall scatter in the derived abundances is $<0.1$~dex
(see ORC02 for a more quantitative comparison).
The molecular oscillator strengths and excitation potentials have been
computed as described in Origlia et al. (1993).
The dissociation energies of the molecules are from Table~2 in
Johnson et al. (1980), the reference Solar abundances from
Grevesse \& Sauval (1998).

Photometric estimates of the stellar parameters are initially used 
as input to produce a grid of model spectra, 
allowing the abundances 
and abundance patterns to vary over a large range and the stellar parameters 
around the photometric values.  
The model which better reproduces the overall observed spectrum and 
the line equivalent widths of selected lines is targetted as the best-fit model.
We measure equivalent widths in the observed spectrum, and in
the best fit model and for four models which are, respectively, 0.1 and
0.2 dex away from the best fit. This approach gives us the uncertainties
listed in Table 2.

Uncertainties in the stellar parameters of $\pm$200~K in temperature (T$_{eff}$), $\pm$0.5~dex in 
log-gravity (log~g) and $\pm$0.5~km~s$^{-1}$ in microturbulence velocity ($\xi$), 
can introduce a further systematic $\le$0.2~dex uncertainty in the 
absolute abundances.
However, since the CO and OH molecular line profiles are very sensitive to 
effective temperature, gravity, and microturbulence variations, 
they constrain better the values of these parameters,  
significantly reducing their initial range of variation and
ensuring a good self-consistency of the overall spectral
synthesis procedure (see ORC02).
Solutions with  
$\Delta $T$_{\rm eff}$$=\pm$200~K, $\Delta $log~g=$\pm$0.5~dex and
$\Delta \xi$$=\mp$0.5~km~s$^{-1}$ and corresponding 
$\pm$0.2~dex abundance variations from the best-fit one are indeed less statistically 
significant (typically at $1\le\sigma\le3$ level only, see Sect.~4). 
Moreover, since the
stellar features under consideration show a similar trend
with variations in the stellar parameters, although with different
sensitivities, {\it relative } abundances are less
dependent on stellar parameter assumptions and their values can be
well constrained down to $\approx\pm$0.1~dex.

\section{Results}

By combining full spectral synthesis analysis with equivalent width
measurements
we derive accurate abundances of Fe, C, O
and [O/Fe] and $^{12}$C/$^{13}$C abundance patterns for the four observed
giants in Terzan~4 and Terzan~5.
The abundances of other $\alpha-$elements, Ca, Si, Mg and Ti, are obtained
by measuring a few strong neutral lines.
These lines are generally heavily saturated and are consequently somewhat
less sensitive to abundance
variations than the molecular CO and OH lines,
which give a very accurate C and O abundances
(see ORC02).

From the NIRSPEC spectra we also derived their heliocentric radial velocities
with a $\pm$5~km~s$^{-1}$ uncertainty.
Average values of
$-50\pm3$~km/s with a dispersion $\sigma\pm5.5$~km/s in Terzan~4
and of
$-93\pm2$~km/s with a dispersion $\sigma\pm3.6$~km/s in Terzan~5
have been obtained.
The inferred radial velocities are fully consistent with those listed by Harris
(1996;1999),
namely --53 and --94~km~s$^{-1}$ for Terzan~4 and Terzan~5, respectively.

Table~1 lists the (V--I)$_0$ colors, heliocentric radial velocity and
the measured equivalent widths of a few representative lines.
Table~2 lists the stellar parameters adopted in the spectral synthesis analysis
and the derived element abundances for all the program stars.
An overall [$\alpha$/Fe] abundance ratio, as the average of
the single [Ca/Fe], [Si/Fe], [Mg/Fe] and [Ti/Fe]
abundance ratios has also been computed.

Stellar temperatures are both estimated from colors and
molecular lines,
gravity from theoretical evolutionary tracks,
according to the location of the stars on the Red Giant Branch (RGB), and adopting
an average microturbulent velocity 2.0 km/s
(see also Origlia et al. 1997).
Equivalent widths are computed by Gaussian fitting the line profiles and
the overall uncertainty is $\le$10\%.

Given the high reddening and crowding in the core region of
these clusters, infrared photometry should be more suitable
to infer reliable stellar temperatures and bolometric luminosities.
Unfortunately, no J,H,K photometry
of the brightest portion of the RGB in the central region of these
clusters is currently available.

\subsection{Terzan~4}

In order to obtain a first guess estimate of the stellar temperatures,
we use the existing $V, I$ photometry of Ortolani,
Barbuy, \& Bica (1997), and we adopt their E(B-V)=2.35 and distance
of 8.3 kpc.  We also use the color-temperature transformation and
bolometric corrections of Montegriffo et al. (1998), specifically
calibrated for globular cluster giants.
We find effective temperatures between 4750 and 5000~K, and
$M_{bol}=-3.5$ to $-4.0$.
Ortolani et al. (1997) also suggests
an overall metallicity as low as [Fe/H]=--2.0.
According to the empirical calibrations of the RGB
features by Ferraro et al. (2000),
at such a low metallicity
the RGB Tip is expected to be located at
M$_{bol}\approx -3.5$ and T$_{eff}\approx$4200~K,
that is at a rather lower temperature than the one estimated
by the (V--I)$_0$ reddening corrected color.

However, one has to take into account that Terzan~4
is a relatively low luminosity
cluster with high degree of field contamination.
The (V,V-I) CMD
in the central region of Terzan~4
obtained by Ortolani, Barbuy, \& Bica 1997 (see their Figure~4)
and used to estimate the cluster reddening and distance
is indeed very sparse and poorly populated.
The barely defined RGB and HB sequences can be equally fitted
with globular cluster mean ridge lines with quite different metallicities in
the range
$\rm -2.0\lesssim [Fe/H]\lesssim -1.0$, by varying the reddening
correction between $\rm 2.3\lesssim E(B-V)\lesssim 1.8$.
Moreover, despite the very good seeing, by inspecting the V and I images
and by computing Point Spread Function photometry
(as also done by Ortolani, Barbuy, \& Bica 1997)
severe blending is still affecting the stars in the center,
particularly star \#2 and \#3 in our sample,
whose luminosities are very likely overestimated.

From our overall spectral analysis and in particular by simultaneously best-fitting
the molecular line intensity and profiles
we find much lower stellar temperatures of $\approx$3800-4000~K
for the observed stars.
By adopting these spectroscopic temperatures an overall [Fe/H]$=-1.60$
and a substantial $\alpha$ enhancement by a factor of $\approx$3 is obtained
(see Table~2).
We also measure a modest carbon depletion ([C/Fe]=--0.25$\pm0.14$~dex)
and low ($\le$10) $^{12}$C/$^{13}$C
abundance ratios.

Fig.~2 shows
our synthetic best fits superimposed on the
observed spectra of the four giants in Terzan~4
in three major spectral regions of interest.
Fig.~3 shows a small portion of the stellar spectra centered on the
$^{12}$CO(6,3) bandhead and on a couple of OH molecular
lines around 1.62 $\mu$m,
which are very sensitive to the stellar temperature.
Superimposed on the observed spectra are two synthetic spectra at [Fe/H]=--1.60
adopting both the photometric and spectroscopic estimates of the
stellar temperatures.
The warmer spectra are too shallow to account for the observed
molecular features since at temperatures significantly above 4000~K
molecules barely survive.

The CO and in particular the OH molecular bands are
indeed extremely sensitive thermometers
in cool giants and make it possible to overcome most of the
problems and uncertainties with
photometric estimates of stellar temperatures in heavely reddened
environments.

In order to check further the statistical significance of our best-fit solution, we compute
synthetic spectra with 
$\Delta $T$_{\rm eff}$$=\pm$200~K, $\Delta $log~g=$\pm$0.5~dex and
$\Delta \xi$$=\mp$0.5~km~s$^{-1}$, and with corresponding simultaneous variations 
of $\pm$0.2~dex of the C and O abundances to reproduce the depth of the molecular 
features.
As a figure of merit we adopt
the difference between the model and the observed spectrum (hereafter $\delta$).
In order to quantify systematic discrepancies, this parameter is
more powerful than the classical $\chi ^2$ test, which is instead
equally sensitive to {\em random} and {\em systematic} scatters
(see also Origlia et al. 2003).
 
Since $\delta$ is expected to follow a Gaussian distribution,
we compute $\overline{\delta}$ and the corresponding standard deviation
for the best-fit solution and the {\it test models}
with the stellar parameter and abundance variations quoted above.
We then extract 10,000 random subsamples from each
{\it test model} (assuming a Gaussian distribution)
and we compute the probability $P$
that a random realization of the data-points around
a {\it test model} display a $\overline{\delta}$ that is compatible
with the {\em best-fit} model.
$P\simeq 1$ indicates that the model is a good representation of the
observed spectrum.

The statistical test has been separately performed on portions of the spectrum
mainly containing the CO bandheads or the OH lines.
Fig.~4 shows the average results for the four stars measured in Terzan~4
with varying T$_{\rm eff}$, log~g and $\xi$, respectively.
As can be appreciated our best fit solution  
gives in all cases a clear maximum in $P$ ($>$99\%)
with respect to the {\it test models}.
OH lines are very efficient in constraining the stellar temperature, while 
CO features are more sensitive to microturbulence:
models with $\Delta $T$_{\rm eff}$$=\pm$200~K and 
$\Delta \xi$$=\pm$0.5~km~s$^{-1}$, respectively, are only 
significant at $> 1.5\sigma$ level.
CO features and to somewhat less extent the OH lines are also sensitive to 
gravity:  
models with $\Delta$log~g$=\pm$0.5~km~s$^{-1}$
lie at $\sigma \gtrsim 1$ from the best-fit solution.
 
The cooler stellar temperatures inferred by our spectral analysis,
which are also in better agreement with the empirical calibrations of the tip RGB
temperatures by Ferraro et al. (2000),
require a somewhat lower reddening correction, namely E(B--V)$\le$2.0,
but near infrared photometry would be most valuable to measure better 
the true reddening and distance of this cluster.

Given the large uncertainty affecting the photometric estimate of
the stellar temperatures in this cluster and as a further check
of the reliability of our overall spectral analysis, we also perform the following
experiment.   Acceptable fits to  the observed H-band 
spectra have been computed
by adopting different temperatures in the range $\rm 3800 \le T_{eff} \le 5000$.
The corresponding range in variation of the iron abundance turns out to be rather small,
between $\approx$--1.6  and --1.4,
while more dramatic variations of the carbon and oxygen abundances have to be adopted.
Figure~5 shows the average [O/Fe] and [C/Fe] abundance ratios required to
fit the observed spectra.
For temperatures above $\approx$4500~K, [C/Fe] exceeds the solar value,
while normally in RGB stars it is somewhat depleted due to the first dredge-up.
More dramatic, for temperatures above $\approx$4200~K, [O/Fe] abundance ratios exceeding
10 times the solar value and up to $\approx$150\ at $\rm T_{eff}\approx5000$~K
are needed to fit the relatively deep OH features in the observed spectra
(see Figure~3). This is definitely unlikely. Also, large oxygen abunandances
are fully inconsistent with the presence of a blue HB, as inferred from
the CMD, typical of metal poor clusters (see also Ortolani et al. 2001).

\subsection{Terzan~5}
\label{ter5}

We use the V,I photometry of Ortolani, Barbuy
\& Bica (1996), the distance estimate of 5.6 kpc and E(B--V)=2.39 (Barbuy et al. 1998)
and the color-temperature transformation of Montegriffo et al. (1998)
to infer temperatures ranging from 3800 to 4000~K for our four observed
giants (cf. Table~1).
More recently, Cohn et al. (2002) by means of HST NICMOS photometry inferred
a slightly lower E(B-V)=2.16 reddening correction, which implies slightly lower
(by $\approx$200~K) stellar temperatures,
in better agreement with the spectroscopic temperatures
as inferred from the OH lines and reported in Table~2.

According to its RGB morphology (see also Ortolani et al. 2001),
which shows a strong curvature in the optical CMD,
Terzan~5 should be extremely metal rich, close to Solar.
In such a high metallicity regime cool giants near the tip of the RGB suffer
dramatic molecular blanketing effects in the V,I photometric bands,
making it impossible to estimate reliable BC$_V$ and BC$_I$
bolometric corrections.

Best-fitting solutions to the observed spectra of the four
giants in Terzan~5 (see Fig.~6)
give an overall [Fe/H]$\approx-0.2$~dex,
an $\rm [\alpha/Fe]$ enhancement by $\approx$+0.3~dex
and a [C/Fe] depletion $-0.32\pm0.18$~dex
(see Table~2).
A low $^{12}$C/$^{13}$C$\le$10 isotopic abundance ratio has been also measured
in this metal rich cluster.

In order to check further the robustness of our best-fit solution,  
the same statistical test done for Terzan~4 has been repeated here.
Fig.~7 shows the average results of the four stars measured 
in Terzan~5 for the CO and OH features with varying T$_{\rm eff}$, log~g and 
$\xi$, respectively.
Our best fit solution  
provides in all cases a clear maximum in $P$ ($>$97\%)
with respect to the {\it test models}.
Models with $\Delta $T$_{\rm eff}$$=\pm$200~K and 
$\Delta \xi$$=\pm$0.5~km~s$^{-1}$, respectively, are only 
significant at $> 2\sigma$ level,
while models with $\Delta$log~g$=\pm$0.5~km~s$^{-1}$
are still significant at $\approx1\sigma$ level. 

Two additional stars (\#5 and \#6 according to our numbering, or
equivalently \#2189 and \#363, respectively, according to
the nomenclature of Ortolani et al. 1996, see also
Table~3) have been observed in Terzan~5.
These stars have photospheric parameters
similar to stars \#3 and \#4 but slightly lower abundances (by $\approx$0.1~dex).
However, the cluster membership of these additional stars is doubtful
since they show heliocentric radial velocities of
--69 and --47~km~s$^{-1}$, respectively,
significantly lower (by $\approx$24 and 46~km~s$^{-1}$) than
the average cluster value (see Sect.~4).
These stars may plausibly be M giant members of the bulge field population.

\section{Discussion and Conclusions}

Our high resolution spectroscopy confirms the photometric
metallicities of Terzan~4 and Terzan~5 as obtained from the morphology
of the RGB (Ortolani et al. 2001).  These two bulge globular clusters were
thought to lie at the extremes of the bulge abundance distribution; our
detailed abundance analysis indicates that this is indeed the case.
Our findings strengthen the analogy between the properties of the
bulge globular clusters and the stellar field population in the
bulge and remain consistent with a common formation history.

In addition, the detailed study of the abundance patterns
indicates oxygen and other $\alpha-$element enhancement at the level
of $\approx+0.5$~dex in the metal--poor cluster, typical of the halo
population, and of $\approx +0.3$~dex in
the metal rich one.
The enhanced $\alpha-$element abundances for Terzan~5 are noteworthy,
as enhanced alphas at Solar metallicity is a hallmark of bulge
stars, reinforcing the scenario of a rapid enrichment of the bulge, similar
to the halo but possibly requiring a higher rate of star formation
(see e.g. Matteucci, Romano \& Molaro 1999; Wyse 2000).

The direct measurement of the $\rm ^{12}C/^{13}C$ abundance ratio
provides major clues to the efficiency of the mixing processes
in the stellar interiors during the evolution along the RGB.
Both the metal-poor and metal-rich giants in the
bulge clusters show very low  $^{12}$C/$^{13}$C$\le10$ ratios.
The classical theory (Iben 1967; Charbonnel 1994 and reference therein)
predicts a decrease of $^{12}$C/$^{13}$C$\approx40$ after
the first dredge--up, the exact amount mainly depending on the chemical
composition and the extent of the convective zone.
However, much lower $^{12}$C/$^{13}$C$\le$10 values have been
measured in several metal-poor halo giants both in the field and in
globular clusters (see e.g. Suntzeff \& Smith 1991, Shetrone 1996,
Gratton et al. 2000 and references therein).
Also the bright giants of $\omega$~Centauri show
very low $^{12}$C/$^{13}$C$\le$7 over the whole metallicity range spanned
by the multiple stellar population in the cluster
(see Brown \& Wallerstein 1993; Zucker, Wallerstein \& Brown 1996;
Wallerstein \& Gonzalez 1996; Vanture, Wallerstein \& Suntzeff 2002;
Smith et al. 2002), including the most metal-rich RGB stars (see
Origlia et al. 2003).
Additional mixing mechanisms due to further {\it cool bottom processing}
(see e.g. Charbonnel 1995; Denissenkov \& Weiss 1996;
Cavallo, Sweigart \& Bell 1998; Boothroyd \& Sackmann 1999;
Weiss, Denissenkov \& Charbonnel 2000)
are thus a common feature during the evolution along the RGB,
regardless the stellar metallicity and environment.

\acknowledgments

LO acknowledges the financial support by the Agenzia Spa\-zia\-le Ita\-lia\-na
(ASI) and the Ministero dell'Istru\-zio\-ne, Universit\`a e Ricerca (MIUR).
RMR acknowledges support from grant number AST-0098739,
from the National Science Foundation.
The authors are grateful to the staff
at the Keck observatory and to Ian McLean
and the NIRSPEC team.
The authors wish to recognize and acknowledge the very significant cultural
role and reverence that the summit of Mauna Kea has always had within
the indigenous Hawaiian community.
We are most fortunate to have the opportunity to conduct observations 
from this mountain.

\clearpage


\begin{deluxetable}{lcccccccccccc}
\footnotesize
\tablewidth{16.5truecm}
\tablecaption{
(V--I)$_0$ colors, heliocentric radial velocity and
equivalent widths (m\AA) of some representative lines
for the observed stars in Terzan~4 and Terzan~5.}
\tablehead{
\colhead{}&
\colhead{}&
\colhead{Terzan~4}&
\colhead{}&
\colhead{}&
\colhead{}&
\colhead{}&
\colhead{}&
\colhead{}&
\colhead{Terzan~5}&
\colhead{}&
\colhead{}&
\colhead{}
}
\startdata
star                & \#1 & \#2 & \#3& \#4 &&&\#1  & \#2 & \#3  & \#4 & \#5 & \#6\\
ref \#$^a$          & 471 &510  & 456& 464 &&& 1614&2167 & 3235 & 401 &2189 &363\\
(V--I)$_0^b$        &1.55 &1.49 &1.34& 1.42&&& 1.81&1.81 & 1.97 &2.10& 2.15 &2.52\\
$v_r$ [km~s$^{-1}$] &--55 &--54 &--45& --45&&&--88 &--94 & --95 &--96 & --69 & --47\\
Ca~$\lambda $1.61508& 49  &60   &  26&   30&&& 208 & 262 &  287 & 297 & 212 & 234\\
Fe~$\lambda $1.61532& 73  &72   &  57&   57&&& 229 & 254 &  270 & 262 &249 &249\\
Fe~$\lambda $1.55317& 51  &57   &  50&   50&&& 215 & 223 &  213 & 200 &178 & 187\\
Mg~$\lambda $1.57658& --  &--   &--  &   --&&& 445 & 447 &  437 & 438 & --  & 428\\
Mg~$\lambda $1.59546& 49  &55   &  50&   44&&& --  & --  &   -- &  -- & 188 & --\\
Si~$\lambda $1.58884& 386 &416  & 361&  377&&& 505 & 518 &  534 & 530 & 481 & 495\\
OH~$\lambda $1.55688& 106 &212  &  87&  106&&& 307 & 334 &  355 & 358 & 347 & 344\\
OH~$\lambda $1.55721& 100 &217  &  78&  109&&& 295 & 338 &  365 & 367 & 357 & 355\\
Ti~$\lambda $1.55437& 236 &246  & 196&  196&&& 404 & 443 &  450 & 429 & 421 & 413\\
\enddata
\tablecomments{
$ $\\
$^a$ Stars in Terzan~4 from Ortolani et al. (1997),
in Terzan~5 from Ortolani et al. (1996).\\
$^b$ Reddening corrected colors adopting E(B-V)=2.0 for Terzan~4 and
E(B-V)=2.16 for Terzan~5 (see Sect.~4).
}
\end{deluxetable}

\begin{deluxetable}{lccccccccccc}
\footnotesize
\tablewidth{17.2truecm}
\tablecaption{Adopted stellar atmosphere parameters and abundance estimates.}
\tablehead{
\colhead{}&
\colhead{}&
\colhead{Terzan~4}&
\colhead{}&
\colhead{}&
\colhead{}&
\colhead{}&
\colhead{}&
\colhead{Terzan~5}&
\colhead{}&
\colhead{}&
\colhead{}
}
\startdata
star           		 & \#1  & \#2  & \#3  & \#4  &&\#1  & \#2  & \#3  & \#4 & \#5 & \#6  \\
T$_{\rm eff}^a$ [K]      & 3800 & 3800 &4000  &4000  && 4000 & 3800 & 3600 & 3600 &3600 &3600\\
log~g                    & 0.5  & 0.5  &1.0   &1.0   && 1.0  & 0.5  & 0.5  & 0.5 & 0.5 &0.5\\
$\xi$ [km~s$^{-1}$]      & 2.0  & 2.0  & 2.0  & 2.0  && 2.0  & 2.0  & 2.0  & 2.0 & 2.0 & 2.0 \\
$\rm [Fe/H]$             &--1.60&--1.58&--1.64&--1.60&&--0.22&--0.19&--0.23&--0.20& --0.33&--0.30\\
                         &$\pm$.10 &$\pm$.10 &$\pm$.11 &$\pm$.11 &&$\pm$.09 &$\pm$.10 &$\pm$.07 &$\pm$.10 &$\pm$.08 &$\pm$.09\\
$\rm [O/Fe]$             & +0.40& +0.60& +0.55& +0.62&& +0.31& +0.30& +0.28& +0.25&+0.29 &+0.26\\
                         &$\pm$.11 &$\pm$.11 &$\pm$.13 &$\pm$.13 &&$\pm$.11 &$\pm$.11 &$\pm$.11 &$\pm$.12 &$\pm$.11 &$\pm$.12\\
$\rm [Ca/Fe]$             & +0.53 & +0.58& +0.54& +0.50&& +0.32& +0.29& +0.33& +0.30&+0.33&+0.30\\
                         &$\pm$.16 &$\pm$.18 &$\pm$.19 &$\pm$.19 &&$\pm$.14 &$\pm$.12 &$\pm$.11 &$\pm$.13 &$\pm$.12 &$\pm$.12\\
$\rm [Si/Fe]$             & +0.57& +0.55& +0.59& +0.50&& +0.37& +0.29& +0.33& +0.40&+0.38&+0.30\\
                         &$\pm$.15 &$\pm$.15 &$\pm$.17 &$\pm$.19 &&$\pm$.18 &$\pm$.18 &$\pm$.17 &$\pm$.18 &$\pm$.15 &$\pm$.15\\
$\rm [Mg/Fe]$             & +0.39& +0.43& +0.42& +0.40&& +0.32& +0.22& +0.28& +0.30&+0.38&+0.30\\
                         &$\pm$.18 &$\pm$.19 &$\pm$.19 &$\pm$.19 &&$\pm$.15 &$\pm$.15 &$\pm$.14 &$\pm$.16 &$\pm$.17 &$\pm$.18\\
$\rm [Ti/Fe]$             & +0.40& +0.47& +0.44& +0.44&& +0.37& +0.34& +0.35& +0.25&+0.33&+0.20\\
                         &$\pm$.19 &$\pm$.19 &$\pm$.19 &$\pm$.18 &&$\pm$.14 &$\pm$.14 &$\pm$.14 &$\pm$.15 &$\pm$.13 &$\pm$.14\\
$\rm [\alpha/Fe]^a$       & +0.47& +0.51& +0.50& +0.46&& +0.34& +0.28& +0.32& +0.31& +0.35&+0.27\\
                         &$\pm$.13 &$\pm$.13 &$\pm$.14 &$\pm$.14 &&$\pm$.12 &$\pm$.12 &$\pm$.10 &$\pm$.12 &$\pm$.10 &$\pm$.11\\
$\rm [C/Fe]$             &--0.40&--0.22&--0.06&--0.30&&--0.48&--0.31&--0.07&--0.40& --0.47&--0.30\\
                         &$\pm$.08 &$\pm$.05 &$\pm$.06 &$\pm$.05 &&$\pm$.05 &$\pm$.07 &$\pm$.05 &$\pm$.04 &$\pm$.06 &$\pm$.05\\
\enddata
\tablecomments{
$ $\\
$^a$ $\rm [\alpha/Fe]$ is the average $\rm [<Ca,Si,Mg,Ti>/Fe]$ abundance ratio
(see Sect.~4).
}
\end{deluxetable}

\clearpage
\begin{figure}
\epsscale{1.0}
\plotone{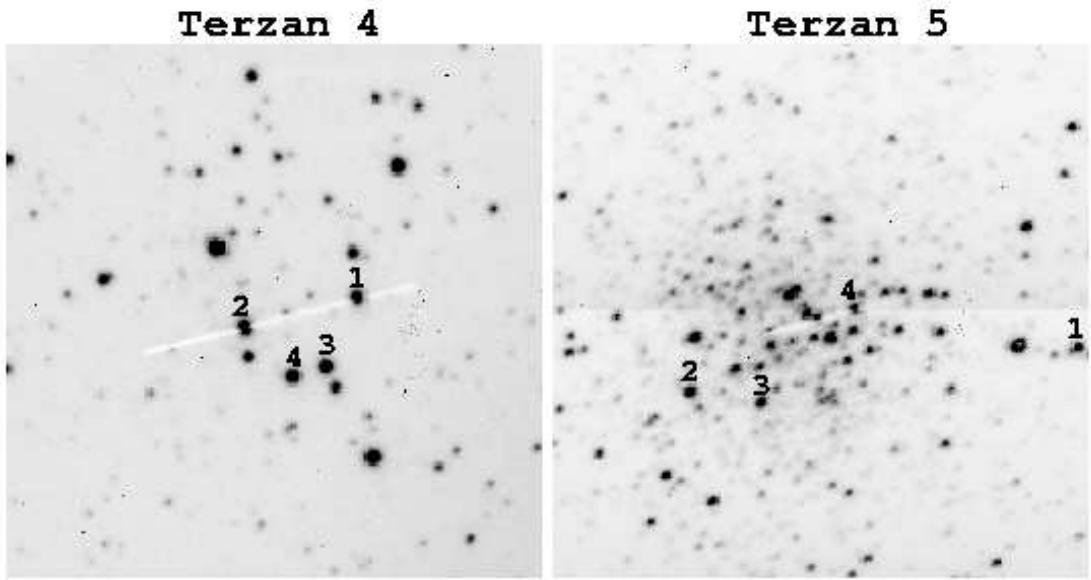}
\caption{
H band images of the core regions of Terzan~4 and Terzan~5 as
imaged by the slit viewing camera (SCAM) of NIRSPEC.
The field of view is 46\arcsec on a side
and the image scale is
$0\farcs183$$~pixel^{-1}$; the slit is 12\arcsec\ long
the observed stars are numbered (cf. Table~1 and 3).
}
\end{figure}

\clearpage
\begin{figure}
\epsscale{1.0}
\plotone{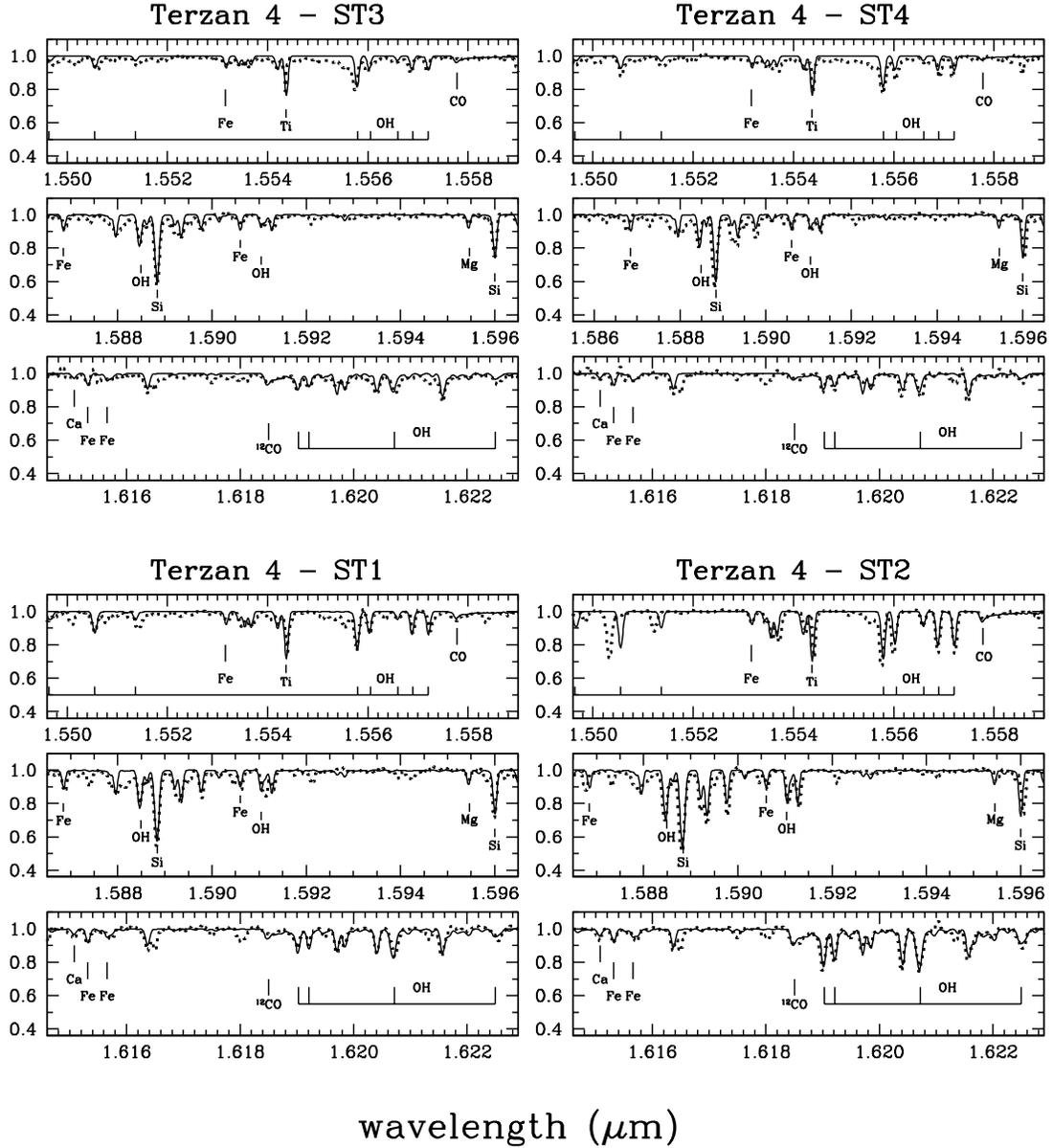}
\caption{
Selected portions of the observed echelle spectra (dotted lines) of the
four giants in Terzan~4 with our best fit synthetic spectrum
(solid line) superimposed. A few important molecular and atomic lines
of interest are marked.
}
\end{figure}

\clearpage
\begin{figure}
\epsscale{1.0}
\plotone{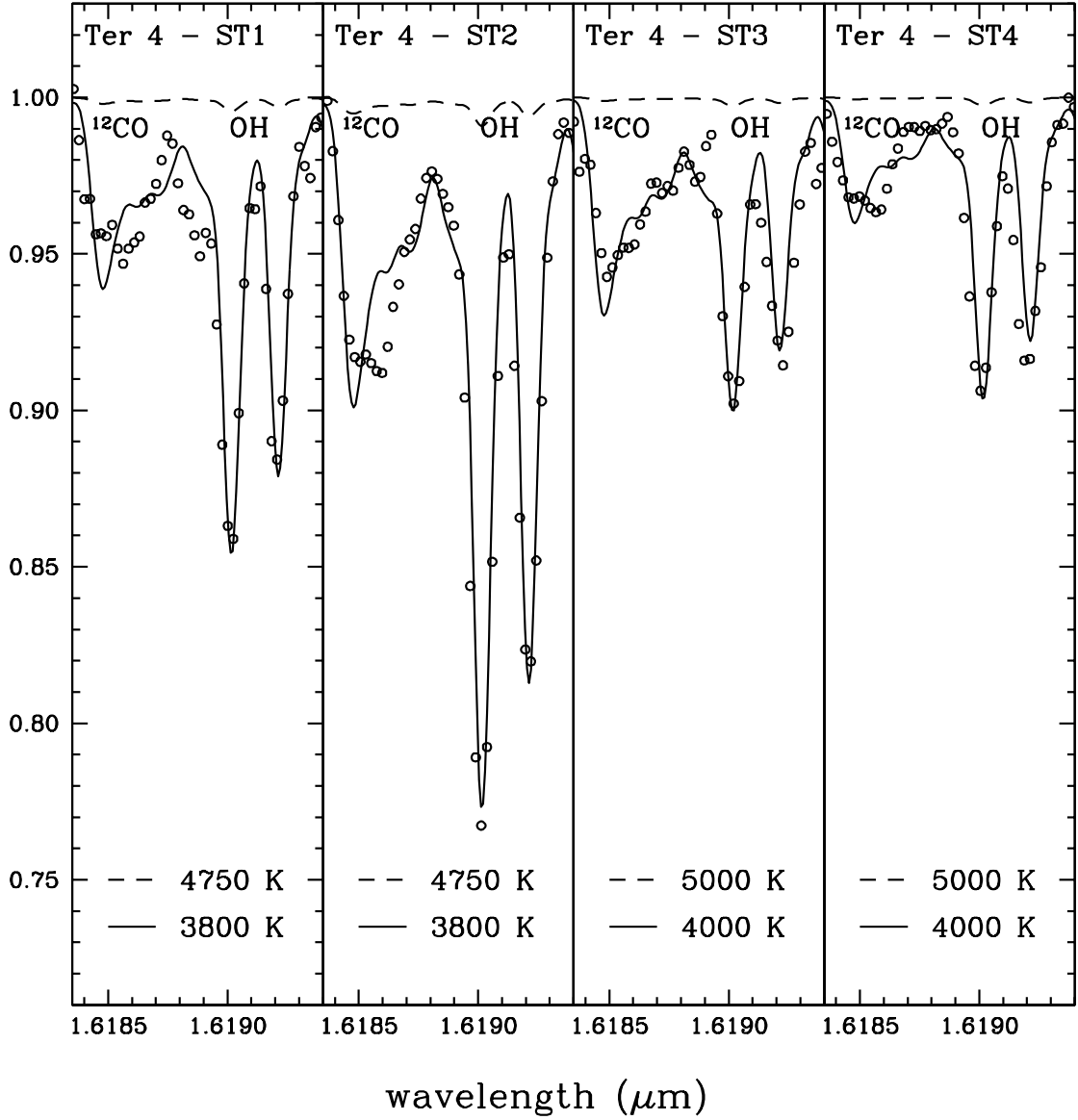}
\caption{
Observed spectra (dotted lines) centered on the
$^{12}$CO(6,3) bandhead and on two OH molecular
lines around 1.62 $\mu$m of the four giants in Terzan~4.
Superimposed are two synthetic spectra at [Fe/H]=--1.6 and
adopting the two different stellar temperatures (dashed and full lines)
reported in the bottom end of each panel (see also Sect.~4).
}
\end{figure}

\clearpage
\begin{figure}
\epsscale{1.0}
\plotone{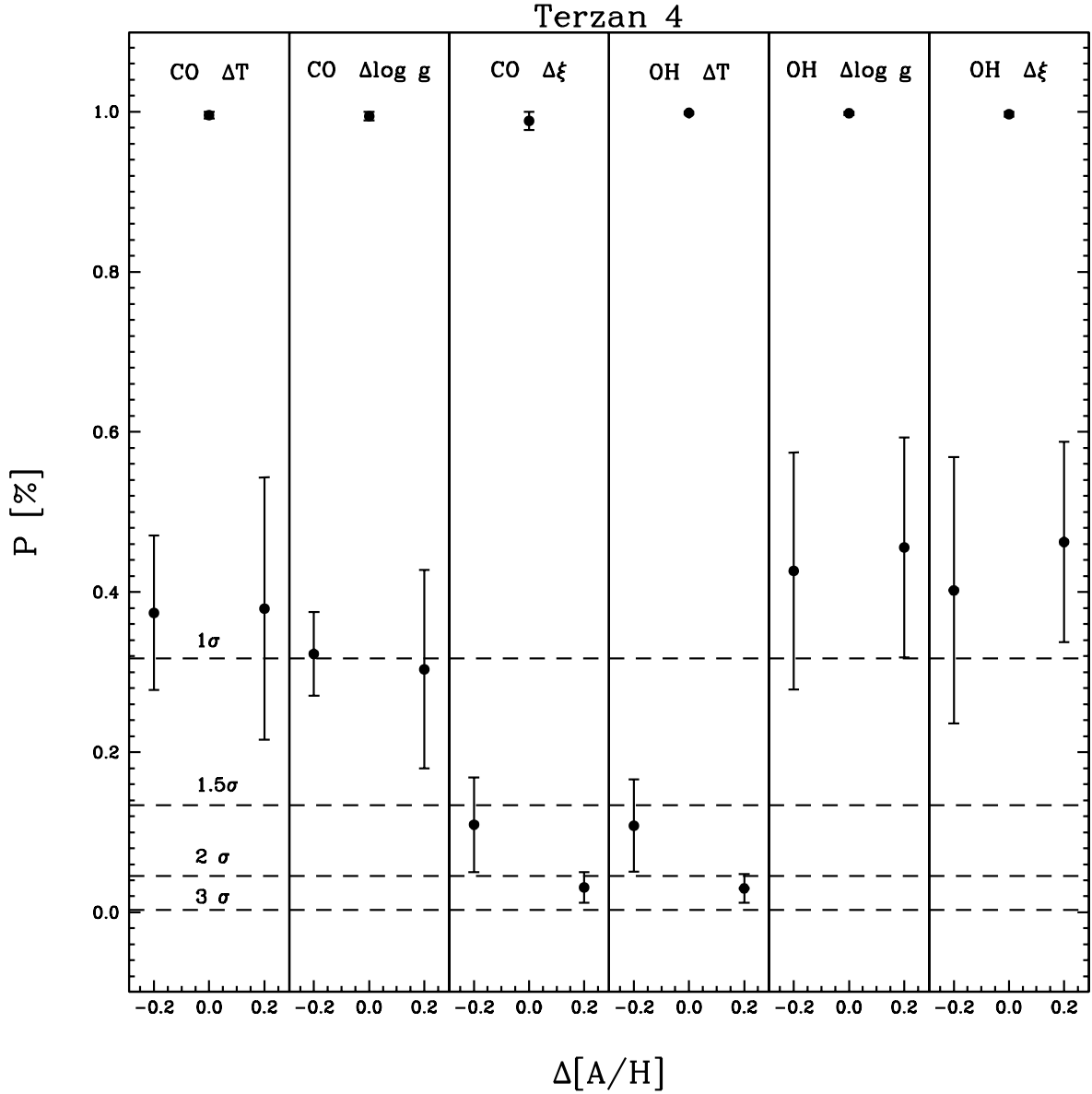}
\caption{ 
Average probability of a Monte Carlo realization for the best
fit solution for the four stars in Terzan~4 (see text).  This
figure shows that our best fit model is stable with respect
to variations in the stellar parameters.
The x-axis is 
The best fit is compared 
with $\Delta[A/H]$ of $\pm$0.2 with respect to
the best-fit to compensate for 
a $\Delta $T$_{\rm eff}$$=\pm$200~K, $\Delta $log~g=$\pm$0.5~dex and
$\Delta \xi$$=\pm$0.5~km~s$^{-1}$ variations, respectively. 
Starting from the left, the first three panels show the results for the 
the CO band-heads, the other three panels for the OH features.
Dashed lines indicate 1, 1.5, 2 and 3$\sigma$ probabilities.} 
\end{figure}

\clearpage
\begin{figure}
\epsscale{1.0}
\plotone{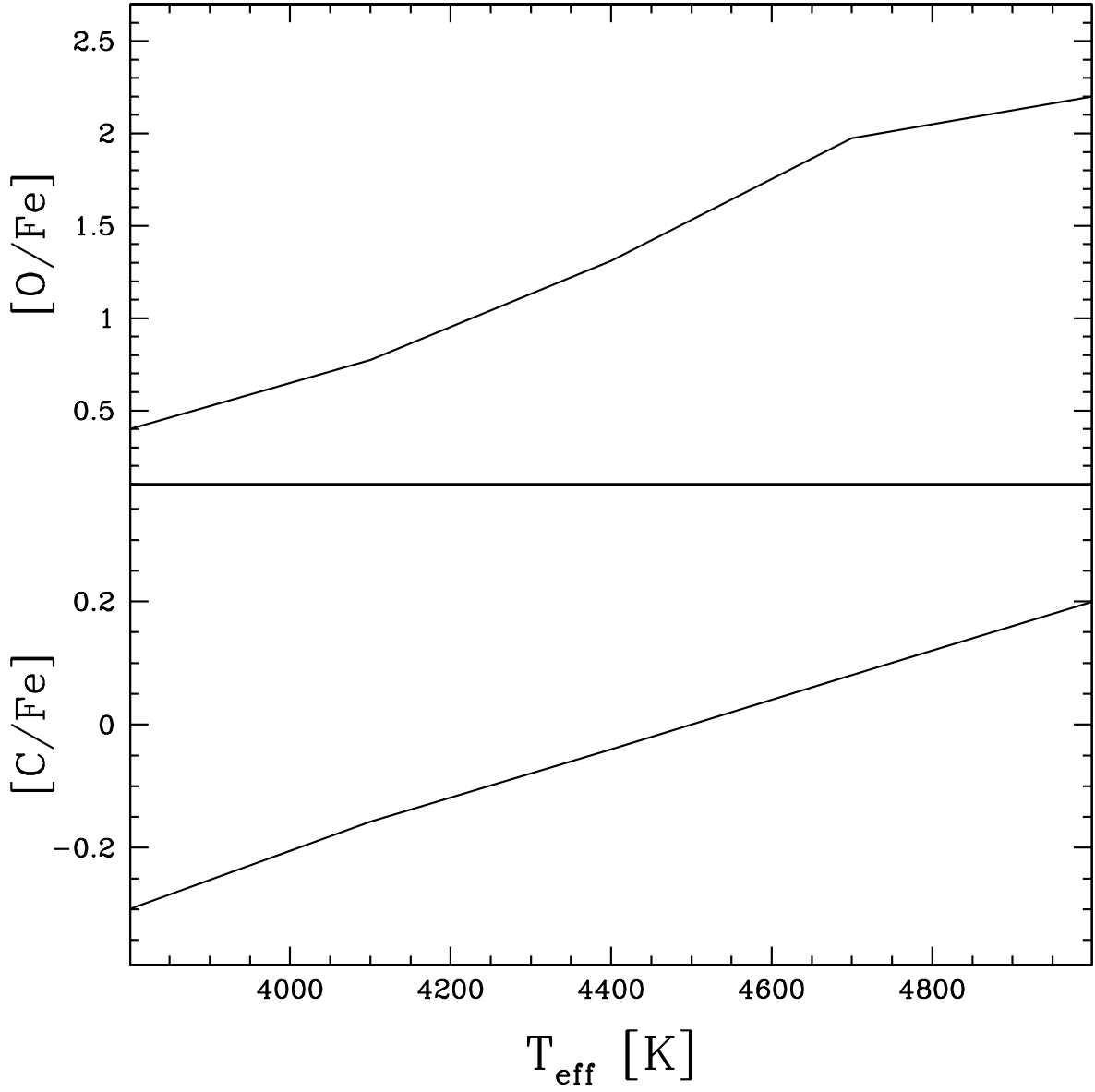}
\caption{
Average [C/Fe] (lower panel)  and [O/Fe] (upper panel) abundance ratios
as inferred by adopting different stellar temperatures for the four giant stars in Terzan~4
(see also Sect.~4).  Notice that adopting temperatures in excess of 4000K leads to 
estimates of [O/Fe] that are unreasonable. }
\end{figure}

\clearpage
\begin{figure}
\epsscale{1.0}
\plotone{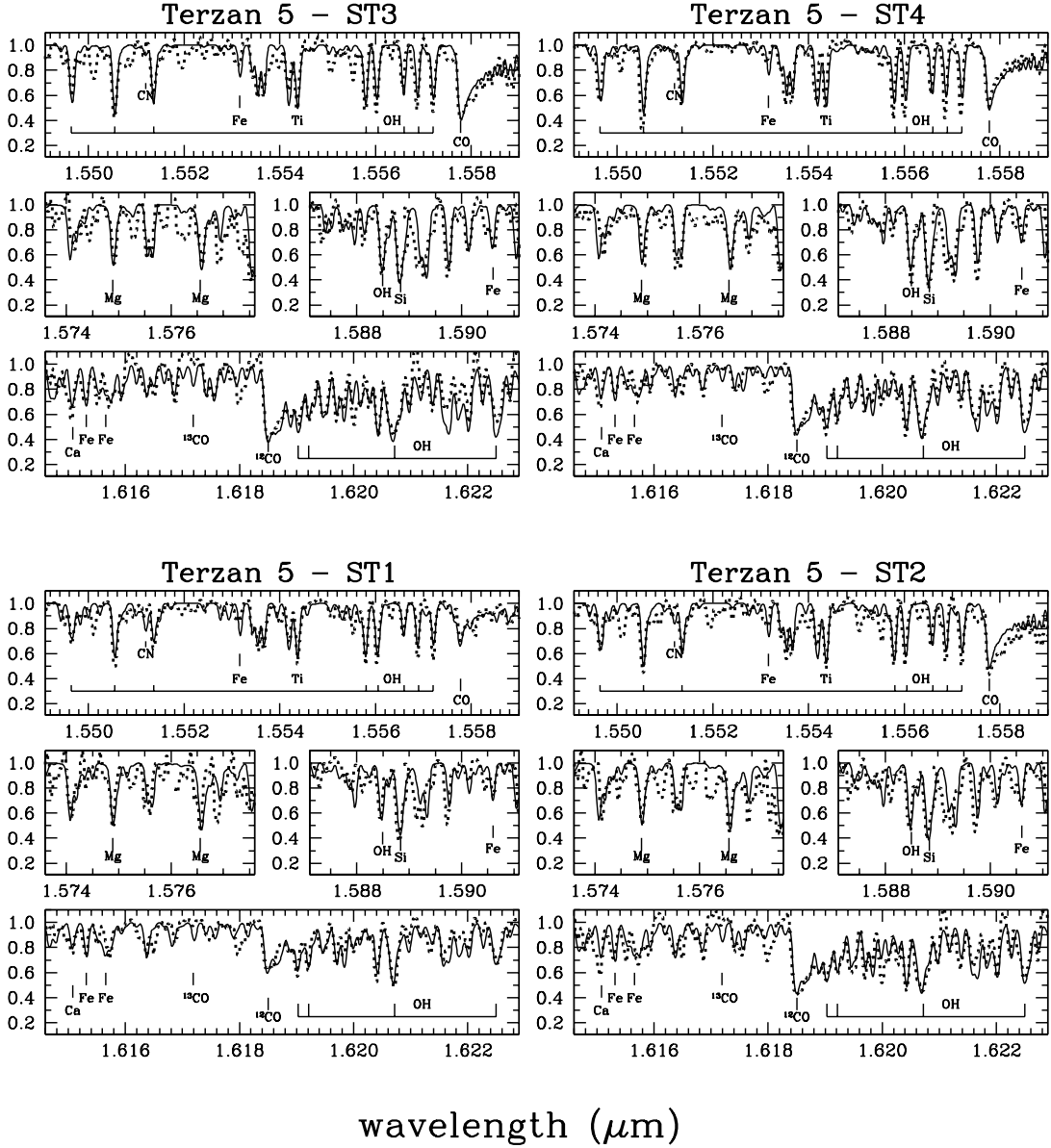}
\caption{
As in Fig.~2 but for the four giant stars in Terzan~5.}
\end{figure}

\clearpage
\begin{figure}
\epsscale{1.0}
\plotone{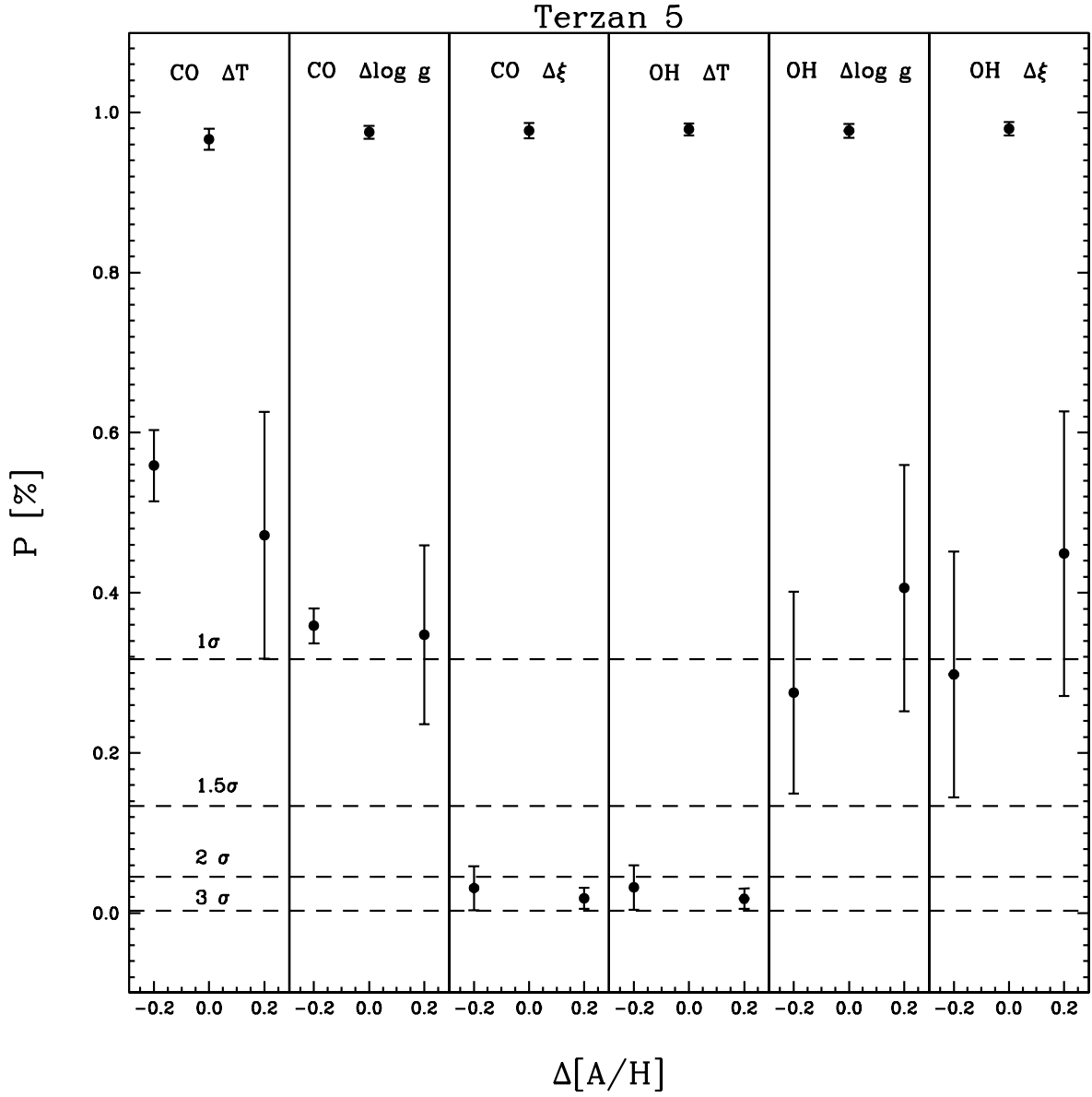}
\caption{
As in Fig.~4 but for Terzan~5.}
\end{figure}

\end{document}